\documentstyle[12pt]{article}

\begin{document}

\title{Comment on ``Nonextensive hamiltonian systems follow Boltzmann's
principle not Tsallis statistics-phase transition, second law of thermodynamics'' by
Gross}

\author{Qiuping A. Wang \\ Institut Sup\'erieur des Mat\'eriaux du Mans, \\
44, Avenue F.A. Bartholdi, 72000 Le Mans, France}

\date{}

\maketitle

\begin{abstract}
Recently, Gross claims that Boltzmann entropy $S=k\ln W$ is valid for any system at equilibrium, so that
Tsallis entropy is useless in this case. I comment on some arguments forwarded to reach this conclusion and
argue that the additive energy formalism of nonextensive statistics is not appropriate for the fundamental
study of the theory for nonadditive systems.
\end{abstract}


\vspace{1cm}

In his recent papers\cite{Gross1,Gross2,Gross3}, Gross wrote ``Boltzmann entropy is well defined,
...independently whether it is extensive or not... the eventual nonextensivity of Hamiltonian systems does not
demand any exotic entropy at equilibrium'', ``there is no alternative to the microcanonical Boltzmann
statistics and to our geometrical foundation of equilibrium statistics'', ``Therefore, for closed Hamiltonian
many-body systems at statistical equilibrium, extensive or not, the thermo-statistical behavior is entirely
controlled by Boltzmann's principle...'' and ``nonextensive Hamiltonian systems do not demand a new entropy
formalism like that by Tsallis''.

Among others, Gross has cited his results of (nonextensive) microcanonical systems of small size to support the
universality of Boltzmann's principle $S=\ln W(E)$ where $W(E)$ is the phase space volume corresponding to
internal energy $E$ (we suppose Boltzmann constant $k=1$). This is certainly a very interesting work and
important affirmation, especially on this moment where the origine and the validity of Tsallis entropy are
still an open question to be discussed from different angles\cite{Beck,Almeida}

One of the important arguments forwarded by Gross to reject Tsallis entropy
$S_q=\frac{W^{1-q}-1}{1-q}$\cite{Tsal88} from the study of nonextensive microcanonical systems {\it at
equilibrium} is the results obtained by Abe\cite{Abe} and Toral\cite{Toral} et al to define thermodynamic
equilibrium for nonextensive systems within, it should be noted, the third version of Tsallis statistics using
escort probability\cite{Penni}. In what follows, I will argue that, as a matter of fact, these results do not
lead to the conclusion against Tsallis entropy for equilibrium nonextensive systems. Let us see this in detail.

It has been shown that\cite{Abe4,Abe3}, for two subsystems $A$ and $B$ of a composite system $A+B$, a
nonadditive entropy
\begin{equation}                                    \label{1}
S_q(A+B)=S_q(A)+S_q(B)+\lambda_SS_q(A)S_q(B)
\end{equation}
and an additive energy $E(A+B)=E(A)+E(B)$ may be a pair of sufficient conditions for the existence of thermal
equilibrium in $A+B$, i.e. $\beta(A)=\beta(B)$ for the inverse temperature\cite{Abe,Toral}
\begin{equation}                                    \label{2}
\beta=\frac{1}{1+(1-q)S_q}\frac{\partial S_q}{\partial E}
\end{equation}
originally defined within the third version of Tsallis statistics. Note that this additive energy is, from the
usual point of view, consistent with the $independent$ joint probability $p(A+B)=p(A)p(B)$ (or
$W(A+B)=W(A)W(B)$ for microcanonical ensemble) for two noninteracting (or weakly interacting, as often
specified) subsystems. This joint probability is essential for Tsallis statistical theory to be $exactly$
applied to $N$-body systems\cite{Wang02b,Wang02c} and, in addition, to get the explicit entropy nonadditivity
Eq.(\ref{1}).

Toral et al \cite{Toral,Velazquez} have shown that the temperature defined in Eq.(\ref{2}) is identical to that
defined in Boltzmann thermo-statistics, i.e. $\beta=\frac{1}{1-q}\frac{\partial \ln [1+(1-q)S_q]}{\partial
E}=\frac{\partial \ln W}{\partial E}=\frac{\partial S}{\partial E}$ for microcanonical ensemble. So it seems
that Tsallis entropy is unnecessary for this case of {\it nonextensive microcanonical systems at equilibrium},
as claimed by Gross.

In view of the situation where the {\it additive energy} formalism of Tsallis statistics is more and more
accepted in the nonextensive statistics, I think that it would be useful to indicate that {\it additive energy}
of {\it noninteracting systems} is perhaps a good approximation for many systems but not appropriate for
discussing fundamental subjects for nonextensive systems implying in general {\it interacting subsystems}. So
that the conclusion of Gross concerning Tsallis theory is only suitable for ``nonextensive systems'' with
additive energy.

Indeed, if the subsystems are independent, one should simply return to additive statistics with, e.g. Boltzmann
or R\'enyi entropy\cite{Reny66}. So it is not surprising to see Tsallis statistics reduced sometimes to $q=1$
due to {\it additive energy}. Toral's interesting result can be even extended (for any version of Tsallis
statistics), as shown by Abe\cite{Abe3}, to canonical ensemble with
$S_q=-\frac{1-\sum_ip_i^q}{1-q}$\cite{Tsal88} satisfying Eq.(\ref{1}), where $p_i$ is the probability that the
system is at the state labelled by $i$ : $\beta=\frac{1}{1-q}\frac{\partial \ln [1+(1-q)S_q]}{\partial
E}=\frac{1}{1-q}\frac{\partial \ln \sum_ip_i^q}{\partial E}=\frac{\partial S^R}{\partial E}$ where
$S^R=\frac{\ln \sum_ip_i^q}{1-q}$ is R\'enyi's entropy\cite{Reny66} which is $additive$ and identical to
Boltzmann entropy for microcanonical ensemble in the case of complete probability distribution
$\sum_{i=1}^wp_i=1$ and $\sum_{i=1}^wp_i^q=w^{1-q}$ where $w$ is the total number of states of the system.

Another example of this self-reduction of Tsallis nonadditive statistics to additive statistics due to additive
energy is the study of nonadditive ideal gas\cite{Abe2} which turns out to be identical to the Boltzmann ideal
gas. A general analysis has been given in the reference \cite{Wang02} where it was shown that, through a series
of theoretical anomalies, the third version of Tsallis statistics might be self-consistent only when $q=1$. In
addition, additive energy destroys Eq.(\ref{1}) for canonical ensemble\cite{Wang02a,Wang02b}.

In conclusion, the discussions of fundamental topics about Tsallis statistics with additive energy are {\it in
general} not appropriate for a nonextensive theory. They cannot be used as arguments against the validity of
Tsallis statistics for equilibrium systems. In my opinion, all the colossal works carried out with additive
energy are very interesting from practical viewpoint, but not good basis for the study of fundamental problems
of nonextensive statistics. It would be helpful for us to return to nonadditive energy, a more general case for
nonextensive systems, within Tsallis statistics and review all the results obtained with additive energy,
especially the derivation of Tsallis entropy from first principles\cite{Almeida}. I would like to indicate here
that the temperature, pressure and the first law of thermodynamics could be well
defined\cite{Wang02c,Wang02,Wang02d} with Tsallis entropy by using the nonadditive energy prescribed by
thermodynamic equilibrium\cite{Wang02a}.

I would like to thank Prof. D.H.E Gross for valuable discussions and for bringing my attention to some recent
papers on the topic. I thank Profs. Toral and S. Abe and L. Velazquez for interesting comments and sending me
references about the topic.

\end{document}